# Observations on recent progress in the field of timing and time perception[1]

**Sundeep Teki**

Auditory Neuroscience Group, Department of Physiology, Anatomy & Genetics,

University of Oxford, South Parks Road, Oxford OX1 3QX, UK.

sundeep.teki@dpag.ox.ac.uk

**Abstract**

Time is an important dimension of brain function, but little is still known about the underlying cognitive principles and neurobiological mechanisms. The field of timing and time perception has witnessed rapid growth and multidisciplinary interest in the recent years with the advent of modern neuroimaging, neurophysiological and optogenetic tools. In this article, I review the literature from the last ten years (2005-2015) using a data mining approach and highlight the most significant empirical as well as review articles based on the number of citations (a minimum of 100 citations). Such analysis provides a unique perspective on the current state-of-the-art in the field and highlights subtopics in the field that have received considerable attention, and those that have not. The objective of the article is to present an objective summary of the current progress in the field of timing and time perception and provide a valuable and accessible resource summarizing the most cited articles for new as well current investigators in the field.

---

[1] This article was originally written for an invited review on Auditory perception and Timing for a special issue of Current Opinion in Behavioural Sciences on 'Interval Timing and Skill Learning: The Multisensory Representation of Time and Action' guest edited by Warren Meck and Richard Ivry. The article was rejected as it did not conform to the specific journal guidelines and is reproduced here in its original form.

## I. Introduction

This is not an editorial piece. For that, please read the editorial article for this special issue on *'Interval Timing and Skill Learning: The Multisensory Representation of Time and Action',* by Richard Ivry and Warren Meck, who are in the best position to comment on the various review articles and provide a succinct summary of the field. My objective is to present my personal observations on the current state of research on timing and time perception, specifically from the point of view of a postdoctoral researcher interested in building an independent research program focused on timing.

My advisor, Tim Griffiths and I were invited by the editor, Warren Meck to contribute a review focused on '*Auditory perception and timing*'. As per the guidelines circulated to all invited authors, the review should focus on, and highlight recent articles of note, especially those published in the last couple of years. My own contribution to the field, during my doctoral research with Tim Griffiths, is represented by two empirical papers (Teki et al., 2011, Teki and Griffiths, 2014) and a review article (Teki et al., 2012) in a special issue on timing, that was also edited by Warren Meck. More recently, Tim and I collaborated with Warren Meck and Melissa Allman on a review elaborating the subjective principles of time perception (Allman et al., 2014).

The guide for authors provided by the Current Opinion in Behavioral Sciences (COBS) clearly stipulates that - "*the aim of the manuscript is to review recent articles, with particular emphasis on those articles published in the **past two years**.*" As the guide encourages the authors to describe recent trends and provide subjective opinions of the topics discussed, I decided to take the liberty to do exactly as instructed, and provide my subjective opinions of recent progress in the field.

A review article is meant to highlight and discuss recent trends and results on a particular topic. Invariably, there is a tendency on the part of most authors to emphasize one's own work at the expense of not adequately discussing results from other research groups. At the outset, I also outlined a possible structure for the review based on my work on duration-based and beat-based timing (Teki et al., 2011, 2012). However, this work does not comply with the publication timeline suggested by the journal. Thus, I decided to not discuss this work which was discussed in a recent review article (Allman et al., 2014). Even if the restricted timeline of two years were to be relaxed, Tim and I struggled to find ten recent articles of note on the particular topic and published in the last two years to emphasize and highlight in our review. After further discussions, we decided to focus on *'Disorders of sequence and interval timing'* based on recent neuropsychological work from Tim's group that examined perceptual timing abilities of patients with striatal and cerebellar degeneration in the form of Huntington's disease, Multiple Systems Atrophy and Parkinson's disease (Cope et al., 2014a, b).

This new topic was more exciting, for the list of authors and topics provided by COBS did not include any clinically oriented articles. Although timing abilities are known to be impaired in a range of neuropsychological and neurological disorders (Allman and Meck, 2012) including Parkinson's disease, Huntington's disease, Multiple Systems Atrophy and Schizophrenia amongst others, the extent to which their timing performance is affected and related to the primary neurological deficits is not fully known. However, even though this field is gaining traction and represents a highly fruitful area of translational research, a review of the recent literature on this topic, again, yielded few promising articles to form the basis of a constructive review that would be beneficial to the field.

At this juncture, it appeared that the two topics of specialization investigated in Tim's lab could not be covered in a detailed and topical review for COBS. At this point, I wondered whether the lack of recent significant breakthroughs is true of the field of timing and time perception as a whole, or is it just a representation of the specific topics we focused on?

**II. Key papers on timing and time perception**

To obtain a representative picture of the field, I decided to examine recent papers by the authors invited to contribute to this special issue (75 authors). As the list of invited authors was a fairly small sample, I extended the list by considering all members of the recently concluded European COST Action - Timely which included members from all over the world. This list provided another 129 authors who were not already invited to contribute to the COBS special issue, thus resulting in a reasonable sample size of 204 authors.

A number of metrics are used to evaluate the quality and impact of research articles including impact factor, h-index, i-10 index amongst others. Although none of these metrics are accepted as standard across the scientific community, I decided on the number of citations as a metric as it indicates the impact of a paper and how well the idea is accepted and circulated in the field by other researchers. It is not a perfect measure, for the number of citations an articles receives is often skewed by the impact factor of the journal where it is published but good ideas tend to circulate well no matter where the ideas are published. Furthermore, to draw reasonable conclusions about progress in the timing field, I focused on a period of 10 years and considered all articles that were indexed in Google Scholar and published by the above set of authors from 2005 onward. In order to restrict my sample of publications and consider only the most impactful papers (ideas), I used a threshold of a minimum of 100 citations. This timeframe is also not ideal, for it is biased towards older papers than more recent articles which have not had the same time to accumulate as many citations.

Using these search criteria, 66 papers (an average of 1 article per every 3 authors) were found as described in Table 1. These papers covered research on topics related to perception of time, rhythm, music, inter-sensory synchrony amongst others and used techniques including psychophysics, neuroimaging, electrophysiology and modeling. Out of these 66 papers, 24 papers were review articles (marked with an asterisk next to the number of citations) that received an average of 223.5 citations (Z-score ranged from - 0.65 to 4.44), i.e. one of out three prominent articles on timing in the last ten years were review articles that discussed the current state of research. The remaining empirical papers, 42 in all, received an average of 181.6 citations per paper (Z-score ranged from - 0.82 to 3.70).

There are several conclusions to be drawn from Table 1, for instance that review articles tend to dominate the overall citations in the field while only an average of four significant empirical papers are published a year. Although many of these reviews are now 'classic' in the field, even the most recent article in the table is a review (Merchant et al., 2013). Among other things, this suggests that either the field is still in an embryonic stage where review articles by established researchers are needed to set the precedent on certain topics or that the field of timing is too diverse, at an intersection of various fields including time perception, rhythm perception, music perception, temporal coding, inter sensory asynchrony, motor timing and coordination, that is reflected in the diversity of topics covered by the review articles. Although not evident from the list, there has been a recent proliferation of review articles given the rise of specialist open access journals like Frontiers that encourage researchers (repeatedly and frequently) to commission special issues covering empirical work as well as reviews. It is not known whether a similar analysis of the most recent and highly cited papers in other fields like memory, vision, or decision-making will yield the same ratio of reviews to empirical studies but one could make a reasonable null hypothesis that this ratio may be smaller than for the field of timing. Alternatively, compared to fields like vision and memory that have been intense topics of investigation for several decades the field of timing is more nascent and does not boast a large research community as evident by the number of participants at specialist meetings in such fields, for instance, the annual Vision Science Society conferences.

In order to drive more experimental work, it is clear that the field of timing needs to attract more young researchers and ensure a a bright future for the field, and this would need concerted efforts from the entire timing community. A recent positive step in this direction is represented by the launch of a specialist journal for timing, *Timing and Time Perception* as well as its corresponding review journal, *Timing and Time Perception Reviews.* Another step forward would be the launch of an academic society exclusively for researchers in timing that would promote interdisciplinary exchange of ideas amongst researchers with diverse interests in timing via annual conferences that draw on a range of methods from purely behavioral to neurophysiological and neuroanatomical measures, and from neurostimulation and neuropsychological approaches to animal work and computational models; share pertinent news and information like grant funding calls, new papers, job opportunities for doctoral and postdoctoral candidates, workshops and training opportunities; and promote career development of young researchers through grants for short cross-disciplinary collaborations or exchange visits, funding for attending conferences and general mentoring support. Although there already exist a few scientific societies and communities relevant to timing like the Society for Music Perception and Cognition (SPMC: http://www.musicperception.org), Rhythm Perception and Production Workshop (RPPW: http://rppw.org), European Society for Cognitive Sciences of Music (ESCOM: http://escom2015.org), Society for Education, Music and Psychology Research (SEMPRE: http://www.sempre.org.uk), Deutsche Gesellschaft fur Musikpsychologie (DGM: http://www.music-psychology.de), Asia-Pacific Society for the Cognitive Sciences of Music, Fondazione Mariani (http://fondazione-mariani.org/) that organizes the NeuroMusic conferences, their scope is limited to music perception and psychology, and do not cover the interests of the entire field of timing and time perception. Society for Neuroscience (SfN) represents the primary venue where researchers gather together for structured mini- or nano-symposia on human and animal timing research but the scientific interaction and discussions are limited given the hectic nature of SfN meetings. A recent example of such a successful academic organization is the Society for the Neurobiology of Language (http://www.neurolang.org/) funded by the National Institutes of Health, which since its inception in 2009, attracts more than 400 researchers for its annual conferences that are held alternatively in the USA (as a satellite meeting of SfN) and Europe.

### III. Future directions

Organizational considerations apart, there are several new scientific directions that the field can and should embrace to achieve a more comprehensive understanding of the neurobiology of timing in natural environments. Animal models of timing focused on core timing networks including the basal ganglia, cerebellum, premotor and parietal cortex (Grahn 2012; Teki et al., 2012; Schneider and Ghose, 2012; Merchant et al., 2013; Allman et al., 2014; Hayashi et al., 2015) will be key to understanding the encoding of time by neuronal ensembles. Such a line of work has been recently pioneered by Hugo Merchant colleagues in rhesus macaques that combines timing behaviors and the examination of the underlying neuronal code in the basal ganglia (Merchant et al., 2011, 2013; Bartolo et al., 2014; Bartolo and Merchant, 2015). Recent work by Mello et al. (2015) further demonstrated that a population code for time exists in the striatum that scales with the interval being timed and multiplexes information about action as well as time. Optogenetic approaches in specific target cells in animal models will yield further crucial insights into the causal role of such mechanisms and their impact on timing behaviors (Grosenick et al., 2015). For instance, a recent study by Chen et al. (2014) reported rapid modulation of striatal activity by the cerebellum via a disynaptic pathway which has implications for the coordinated processing of temporal information in the two core timing areas.

The other dominant view of timing is that it is not a computation of specific dedicated circuits but rather the output of intrinsic neuronal dynamics (Karmarkar and Buonomano, 2007; Ivry and Schlerf, 2008). In this respect, the activity of sensory areas including auditory, visual and somatosensory cortices merits further attention. Combining optogenetics and single-unit recordings in primary visual cortex (V1), Hussain Shuler and colleagues have recently provided beautiful insights into how V1 responses predict and drive the timing of future actions (Namboodiri et al., 2015) and recruit basal forebrain and cholinergic input within V1 to encode the timing of visually cued behaviors (Liu et al., 2015).

In order to obtain a fundamental understanding of timing, it is also imperative to use stimuli and paradigms that mimic timing behaviors in the natural world. Such naturalistic sequences that go beyond the use of single intervals that have been traditionally used will offer additional insights on encoding of time as well as associated motor behaviors (Kornysheva and Diedrichsen, 2014). Table 1 and the reviews therein highlight that timing is not mediated by a single brain area but rather involves a distributed network (Meck, 2005) in cortical and subcortical areas including prefrontal, parietal, premotor and sensory cortices, insula, basal ganglia, cerebellum, inferior olive amongst others. To form a clear picture of how timing is mediated by these structures, it is also important to understand the core functions of these areas and what particular aspect of timing they mediate, whether it is related to attention, memory or perception.

The use of comparative paradigms in healthy human volunteers as well as clinical populations that show timing deficits such as patients with Parkinson's, Huntington's, Schizophrenia amongst others will provide a more uniform understanding of timing functions and dysfunctions in health and disease. An identical approach (and even the use of similar paradigms) in animal models via use of control animals as well as lesion or knock-out models will complement findings from the human literature and provide a more generic understanding of fundamental mechanisms of timing.

Irrespective of the present state of affairs, the field of timing and time perception represents a promising and highly active field of research that is growing every year in terms of number of researchers and scientific output and one where new students and researchers can find a niche topic and leave a significant mark on the field.

**Table** List of 66 papers on timing, rhythm and music perception from 2005 to present sorted according to the number of citations (minimum of 100 citations) in Google Scholar collated during the period from August 30 to September 16, 2015 (see section II for more details). Asterisks next to the number of citations denote review articles.

| No. of citations | Reference | Year | Journal | Summary |
|---|---|---|---|---|
| 1068* | Buhusi & Meck | 2005 | Nat Rev Neurosci | Time is represented in a distributed manner through coincidental activation of cortico-striatal neuronal populations. |
| 550 | Casasanto & Boroditsky | 2008 | Cognition | Spatial information affects judgments about duration but not vice versa. |
| 442 | Wittmann et al. | 2006 | Chronobiol Int | Social jetlag, i.e. the discrepancy between social and biological timing affects wellbeing and stimulant consumption. |
| 417 | Grahn et al. | 2007 | J Cogn Neurosci | Basal ganglia and Supplementary Motor Areas mediate beat perception, in addition to motor production. |
| 310* | Grondin | 2010 | Att Percept Psychophys | Review of recent behavioral and neuroscientific studies of timing. |
| 305 | Karmarkar & Buonomano | 2007 | Neuron | Cortical networks can read out time as a result of intrinsic network dynamics |
| 294* | Ivry & Schlerf | 2008 | Trends Cogn Sci | Dedicated models of timing are preferred over intrinsic models. |
| 292 | Shuler & Bear | 2006 | Science | Primary sensory cortex, like V1, mediates reward-timing activity. |
| 289* | Coull et al | 2011 | Neuropsychopharmacology | Review of neuroimaging, neuropsychological and psychopharmacological aspects of timing. |
| 276 | Morrone et al. | 2005 | Nat Neurosci | Short intervals of time between two successive perisaccadic visual stimuli (but not auditory) are underestimated. |

| No. of citations | Reference | Year | Journal | Summary |
|---|---|---|---|---|
| 274 | Chen et al. | 2008 | Cereb Cortex | Passively listening to rhythms recruits motor regions of the brain. |
| 265* | Droit-Volet & Meck | 2007 | Trends Cogn Sci | Review of how emotional arousal and valence modulates attentional time-sharing and clock speed. |
| 261 | Patel et al. | 2009 | Curr Biol | Snowball, a cuckatoo, can spontaneously synchronize its movements to a musical beat. |
| 253* | Wittman & Paulus | 2008 | Trends Cogn Sci | Review of how impulsivity affects perception of time and decision making. |
| 237 | Winkler et al. | 2009 | Proc Natl Acad Sci | Newborn infants show beat perception. |
| 228 | MacDonald et al. | 2011 | Neuron | Hippocampal time cells encode successive moments during a sequence of events. |
| 222* | Meck | 2005 | Brain & Cognition | Review of timing that suggests a distributed representation of time across multiple neural systems. |
| 220* | Meck et al. | 2008 | Curr Opin Neurobiol | Review that proposes striatum serves as a core timer, as part of a distributed timing system. |
| 219* | Wiener et al. | 2010 | NeuroImage | Meta analysis that suggests distinct for perceptual vs. motor timing; SMA and right IFG are most commonly activated in various timing tasks. |
| 212* | Coull & Nobre | 2008 | Curr Opin Neurobiol | Review that suggests basal ganglia is key for explicit timing while parietal and premotor areas mediate implicit timing. |
| 210* | Nobre et al. | 2007 | Curr Opin Neurobiol | Review that describes how temporal expectations modulate perception and action, and the underlying neural mechanisms. |
| 193 | Patel et al. | 2005 | Brain Res | Beat perception and synchronization show modality specific benefits for auditory vs. visual beat patterns. |

| No. of citations | Reference | Year | Journal | Summary |
|---|---|---|---|---|
| 192 | Chen et al. | 2008 | J Cogn Neurosci | Musicians show greater prefrontal cortex activity vs. non-musicians while tapping to complex auditory rhythms. |
| 177 | Grahn & Rowe | 2009 | J Neurosci | Putamen, SMA and premotor cortex are important for internal generation of the beat and auditory motor coupling during beat perception. |
| 177* | Lewis & Miall | 2006 | Trends Cogn Sci | Dorsolateral prefrontal cortex mediates working memory as well as timing. |
| 175 | Meck | 2006 | Brain Res | Dopamine depleting lesions in different parts of the basal ganglia shows dissociable effects on duration discrimination. |
| 173 | Noesselt et al. | 2007 | J Neurosci | Temporal correspondence between auditory and visual streams modulates activity of multisensory STS as well as unisensory cortices. |
| 170 | Arvaniti | 2009 | Phonetica | Review of work on rhythmic categorization which argues that timing is distinct from rhythm. |
| 168 | Wittmann et al. | 2007 | Exp Brain Res | Posterior insula mediates delayed gratification of reward while striatum encodes time delay. |
| 165* | Kotz & Schwartze | 2010 | Trends Cogn Sci | Review which suggests that temporal and speech processing is processed by cortical and subcortical systems associated with motor control. |
| 162 | Burr et al. | 2007 | Nat Neurosci | Short visual events are encoded by visual neural mechanisms with localized receptive fields rather than by a centralized supramodal clock. |
| 159* | Vroomen & Kreetels | 2010 | Att Percept Psychophys | Review that focuses on intersensory timing and mechanisms that encode intersensory lags. |
| 155* | Wittmann | 2009 | Phil Trans R Soc B | Review that discusses different models of time perception with a particular focus on the insula as a core timer. |

| No. of citations | Reference | Year | Journal | Summary |
|---|---|---|---|---|
| 153 | McAuley et al. | 2006 | J Exp Psychol: General | Event timing profiles for a battery of perceptual-motor timing tasks vary across the life span (4-95 years old). |
| 152 | Chen et al. | 2006 | NeuroImage | Metrical structure of musical rhythms modulates functional connectivity between auditory and dorsal premotor cortex. |
| 151 | Boroditsky et al. | 2011 | Cognition | English and Mandarin speakers think about time differently. |
| 150* | Taatgen et al. | 2007 | Psychol Rev | A time perception model based on adaptive control of thought-rational can explain effects of attention and learning during time estimation. |
| 150* | Correa et al. | 2006 | Brain Res | Review that focuses on how temporal attention modulates the amplitude and latency of ERPs like N2 and P300 components. |
| 144 | Matlock et al. | 2005 | Cogn Sci | Fictive motion influences temporal reasoning. |
| 136* | Allman & Meck | 2012 | Brain | Review that focuses on distortions of time perception and timed performance in various neurological and psychiatric conditions. |
| 135* | Block et al. | 2010 | Acta Psychol | Meta analysis that focuses on the effects of cognitive load on prospective and retrospective duration judgments. |
| 135 | Kotz et al. | 2009 | Cortex | Review that focuses on the non-motor functions of basal ganglia with particular emphasis on prediction in speech and language. |
| 134 | Teki et al. | 2011 | J Neurosci | Perception of relative and absolute time is mediated by distinct networks based in the basal ganglia and the cerebellum respectively. |
| 134 | Lewis & Miall | 2006 | Behav Proc | Dorsolateral prefrontal cortex mediates working memory and posterior parietal cortex and anterior cingulate attentional aspects of timing. |

| No. of citations | Reference | Year | Journal | Summary |
|---|---|---|---|---|
| 130* | Rubia et al. | 2009 | Phil Trans R Soc B | Review that suggests that impulsivity in ADHD is related to compromised timing functions and dopamine dysregulation. |
| 130 | Styns et al. | 2007 | Hum Mov Sci | Walking speed is modulated by the tempo of musical and metronome stimuli. |
| 128 | Kanai et al. | 2006 | J Vis | Temporal frequency of a stimulus serves as the clock for perceived duration. |
| 125 | Noulhiane et al. | 2007 | Emotion | Emotional stimuli are judged longer than neutral stimuli, when balanced for the levels of arousal. |
| 124 | Fuhrman & Boroditsky | 2010 | Cogn Sci | Temporal judgments in nonlinguistic tasks are influenced by culturally specific spatial representations. |
| 122 | Grahn & Brett | 2009 | Cortex | Parkinson's patients show selective deficits in discrimination of beat-based rhythms. |
| 122* | Eagleman et al. | 2005 | J Neurosci | Review of timing based on psychophysics, electrophysiology, imaging and computational modeling. |
| 120 | Keller et al. | 2007 | Consc & Cogn | Action simulation in ensemble musicians like pianists underlies synchronization and self-recognition. |
| 120 | Correa et al. | 2005 | Psychon Bull & Rev | Temporal orienting enhances perceptual processing. |
| 118* | Merchant et al. | 2013 | Ann Rev Neurosci | Review that highlights the role of a core timing mechanism in the basal ganglia and its interaction with context dependent areas. |
| 113 | Grahn & McAuley | 2009 | NeuroImage | Individual differences in beat perception exist and modulate activity in auditory and motor areas. |
| 110 | Zarco et al. | 2009 | J Neurophys | Performance of rhesus monkeys and humans is compared on a number of sub-second interval reproduction tasks. |

| No. of citations | Reference | Year | Journal | Summary |
|---|---|---|---|---|
| 109 | Wearden et al. | 2008 | Q J Exp Psychol | Decreasing arousal affects performance on time perception tasks. |
| 108 | Vatakis & Spence | 2006 | Brain Res | Cross-modal temporal discrimination performance is better for audiovisual stimuli of lower complexity. |
| 107 | Nozaradan et al. | 2011 | J Neurosci | EEG frequency tagging reveals neural entrainment to beat and meter. |
| 107 | Wassenhove et al. | 2008 | PLoS One | Multisensory interactions influence perception of time: vision can impact auditory temporal perception. |
| 107 | Ishihara et al. | 2008 | Cortex | A mental time line exists from left to right along the horizontal axis in space. |
| 105* | Keller | 2008 | Emerg Comm | Review that addresses cognitive processes underlying joint action in music performance. |
| 102 | Casasanto et al. | 2010 | Cogn Sci | Spatial information influences temporal judgments more than time affects spatial judgments in children as well as adults. |
| 101 | Iversen et al. | 2009 | Ann NY Acad Sci | Beta-band activity influences auditory rhythm perception. |
| 100* | Droit-Volet & Gil | 2009 | Phil Trans R Soc B | Review that addresses the role of emotional context on timing. |
| 100 | Jahanshahi et al. | 2006 | J Neurosci | Basal ganglia and cerebellum are involved in reproduction of both short and long intervals. |